\documentclass[12pt]{article}
\title{Nonperturbative mechanisms of  strong decays in QCD}
\author{Yu.A.Simonov\\
 State Research
Center\\Institute of Theoretical and Experimental Physics, \\
Moscow, 117218 Russia}
 \date{}
\newcommand{\beq}{\begin{eqnarray}}
 \newcommand{\eeq}{\end{eqnarray}}
\newcommand{\be}{\begin{equation}}
 \newcommand{\ee}{\end{equation}}

\def\ga{\mathrel{\mathpalette\fun >}}
\def\fun#1#2{\lower3.6pt\vbox{\baselineskip0pt\lineskip.9pt
\ialign{$\mathsurround=0pt#1\hfil ##\hfil$\crcr#2\crcr\sim\crcr}}}

\newcommand{{\SD}}{\rm SD}

\newcommand{\veY}{\mbox{\boldmath${\rm Y}$}}
\newcommand{\vex}{\mbox{\boldmath${\rm x}$}}
\newcommand{\vey}{\mbox{\boldmath${\rm y}$}}
\newcommand{\ver}{\mbox{\boldmath${\rm r}$}}

\begin{document}
\maketitle

\begin{abstract}
Three decay mechanisms are derived systematically from the QCD
Lagrangian using the field correlator method. Resulting operators
contain no arbitrary parameters and depend only on characteristics
of field correlators known from lattice and analytic calculations.
When compared to existing phenomenological models, parameters are
in good agreement with the corresponding fitted values.
\end{abstract}

\section{Introduction}

An enormous amount of experimental data on strong decays of mesons
and baryons is partly used by theoreticians for comparison in the
framework of the $~^3 P_0$ model \cite{1}, and its flux-tube
modifications \cite{2}. The analysis done in \cite{3} confirmed
the general validity of the model, whereas in \cite{4} results of
other forms of decay operators have been also investigated in
meson decays, and in \cite{5} in baryon decays. On the whole, the
phenomenological picture seems to be satisfactory for the $~^3
P_0$ model with some exclusions discussed in \cite{3} and
\cite{6}. The recent extensive study of strong decays of strange
quarkonia  based on the $~^3P_0$ model  was done in \cite{7}.

The key element which is missing in this situation is the
systematic derivation of all terms in the decay Hamiltonian from
the basic principles, i.e. from the QCD Lagrangian. It is the
purpose of the present paper  to make some progress in this
direction using the Field Correlator Method (FCM) \cite{8} and
background perturbation theory \cite{9} to treat nonperturbative
(NP) QCD contributions together with perturbative ones.

In doing so one should take into account the special role of pions
in the hadron decays and therefore to perform accurately the
chiral bosonization of the effective quark Lagrangian, obtained
from the basic QCD Lagrangian. This will give the first term in
the decay Hamiltonian, and the corresponding decay mechanism will
be referred to as a Chiral Decay Mechanism (CDM). At the same time
one should take into account the string degrees of freedom in the
original meson and the possibility of string breaking due to the
$q\bar q$ pair creation. The corresponding term in the decay
Hamiltonian will be derived below without free parameters and this
second mechanism will be called the String Breaking Mechanism
(SBM). As will be seen, the dominant term of SBM has the
structure, which can be compared quantitatively with the
phenomenological fits of the $~^3 P_0$ model in \cite{3, 4}.

Finally, the QCD perturbation theory in the perturbative
background developed in \cite{9}, allows to derive two additional
terms in the decay Hamiltonian: one for the OZI-allowed decays,
which has the Lorentz form of the $~^3 S_1$ type but proceeds
through the intermediate hybrid state, and another for the
OZI-forbidden decays, which proceeds through the intermediate
glueball state and only at very small distances reduces to the
two-gluon or three-gluon $q\bar q$ pair creation. We shall call
these mechanisms the Hybrid Mediated Decay (HMD) and the Glueball
Mediated Decay (GMD) respectively.

\section{The Chiral Bosonization and the Chiral Decay Mechanism}

One starts with the QCD Lagrangian in the Euclidean space-time and
averages over the gluonic fields  writing the general form of the
gauge-invariant correlator (known from lattice or analytic
calculations, see refs. in \cite{8}) which contains the confining
part $D(x)$, namely,
\be
\frac{g^2}{N_c}\langle
tr(F_{\mu\nu}(x)\Phi(x,y)F_{\lambda\sigma}(y)\Phi(y,x)) \rangle =
(\delta_{\mu\lambda}\delta_{\nu\sigma}
-\delta_{\mu\sigma}\delta_{\nu\lambda}) D(x-y) +O(D_1) \label{1}
\ee where $O(D_1)$ contains a relatively small nonconfining part
$D_1(x)$ and $\Phi (x,y) = P\exp ig \int^x_y  A_\mu dz_\mu$ is the
parallel transporter.

Assuming also that all higher correlators can be neglected (as it
is supported by lattice data, see \cite{10}) one obtains the
effectice action \cite{11} $$
 L^{(2)}_{EQL} (el) =\frac{1}{2N_c} \int d^4x\int
 d^4y^f\psi^+_{a\alpha}(x)^f \psi_{b\beta}(x)
 (x)^g\psi^+_{b\gamma} (y)^g\psi_{a\varepsilon} (y)\times
 $$
 \be
 \times [\gamma^{(4)}_{\alpha\beta}
 \gamma^{(4)}_{\gamma\varepsilon} J_E(x,y)+\gamma^{(i)}_{\alpha\beta}\gamma^{(i)}_{\gamma\varepsilon}
 J_M(x,y)]\label{2}
 \ee
where the kernel $J(x,y)$ is expressed through $D(x)$,
 \be
J_{E(M)}(x,y) =\int^x_Y du_i\int^y_Y dv_i K_{xy}^{(E(M))}(u,v)
D(u-v),~~i=1,2,3. \label{3} \ee and $D(x)$ is connected to the
string tension $\sigma$ in the usual way, $\sigma =
\frac{1}{2}\int d^2 x D(x)$. Here the string kernel $J_{E(M)}$
contains color electric (magnetic) fields and the former is
dominant at large distances, and therefore magnetic part will be
omitted for simplicity. At this point one should specify the
choice of integration contours, the kernel $K$ and the point $Y$
in (\ref{3}). As it is known, the full result of the integration
over gluon fields does not depend on the gauge and on the shape of
the contours (when all correlators are taken into account);
however to write the contribution of bilocal correlator (\ref{1})
in the gauge-invariant form, one has to use one of the variants of
the contour gauge, e.g. \cite{12}, where $K_{xy}^{E} \equiv 1$,
and to choose the contour corresponding to the minimal string
which minimizes the contribution of higher correlators. In this
section we consider the following geometry: the quarks at the
points $x, y$ in (\ref{2}) are at one end of the string (they are
dynamically close \cite{11}), while the point $Y$ is at the
position of the heavy antiquark, as it was introduced in
\cite{11}. In the next section we shall consider a more general
geometry, when the points  $Y$ in (\ref{3}) are different and
integration over $du_i$ and $dv_i$ runs over two different  pieces
of the broken string. For the results of the present section the
exact definition of the point $Y$ is inessential since the pion is
emitted from the end of the string under consideration while
another end of string is a spectator.

As the next step the bosonization  of the Lagrangian (\ref{2}) can
be done in the usual way, however with nonlocal mesonic fields
with the result \cite{14}
\be
  \Delta L=i\int d^{4}x d^{4}y\psi^{+}(x)\hat M(x,y)\psi(y)\label{4}
\ee where the kernel $\hat M(x,y)$ can be written as a nonlinear
form for PS meson fields
\be
\hat M(x,y)= M_{S}(x,y)\exp(i\gamma_{5}\hat\phi(x,y)) + ...,~ \hat
\phi=\frac{\phi_A\lambda^A}{F_\pi},~ F_\pi=93{\rm MeV}.
\label{5}\ee Here ellipsis implies all other omitted terms,
including those containing isovector scalars, vector and
pseudovectors.

One should note, that in our case when confinement is present the
constant condensate of scalar-isoscalar field always enters
multiplied by $J(x,y)$ and thus produces   the scalar confining
potential of the string
 in $M_S(x,y)$ \cite{14}. Namely, for long enough string, i.e. for
$|\vex - \veY| \gg T_g$, where $T_g$ is the gluonic correlation
length in $D(x)$, one has approximately \cite{11}
\be
 M_{S}(x,y) \approx \sigma|\vex| \delta^{(4)}(x-y), \label{6}
 \ee
 This is different from the instanton, or the NJL model, where the
 scalar-isoscalar field acquires a nonzero condensate, which is
 constant in all space-time.

 Inserting (\ref{6}) into (\ref{5}) one obtains a localized form
 of the quark-meson Lagrangian, describing interaction on one end
 of the string; to the lowest order in PS field,
\be
 \Delta L^{(1)}= \int \overline{\psi}(x)\sigma |\vex - \veY|\gamma_{5}
 \frac{\phi^{A}\lambda^{A}}{F_{\pi}} \psi(x) dt d^{3}x, \label{7} \ee
 One can visualize in (\ref{7}) the simultaneous presence of quark
 fields $\psi, \bar\psi$, together with the string $\sigma|\vex -
 \veY|$ and Nambu-Goldstone (NG) fields $\phi^A$, $A = 1, ...,
 n_f^2 -1$.

 Using Dirac equation for the quark field $\psi(x)$ one arrives as
 in \cite{14}
 at the familiar Weinberg Lagrangian \cite{15}
\be
 \Delta L^{ch} = g^{q}_{A} tr (\overline\psi\gamma_{\mu} \gamma_{5}
 \omega_{\mu} \psi),\omega_\mu =\frac{i}{2}(u \partial_\mu u^{+}-
 u^{+} \partial_{\mu} u), \label{8} \ee
where $u(x) = \exp(\frac{i}{2}\gamma_5\hat\phi(x,x))$. In our
derivation $g_A^q$ is uniquely defined in the local limit,
\be
g_A^q \equiv 1 \label{9} \ee which agrees with large $N_c$ limit,
discussed in \cite{15}.

Note that both Lagrangians (\ref{7}), (\ref{8}) are local limits
of nonlocal expression (\ref{4}), and for not very long string
with the length $L \sim T_g \sim 0.2$ fm the nonlocality is
essential. However string is apparently not present in (\ref{8}),
both quark operators there are solutions of Dirac equation with
the string entering as a scalar potential. It is clear that the
Lagrangian (\ref{8}) describes the pion field emission both from
the quark at one end of the string and from the antiquark at
another end of the string. In case of baryons one should sum up in
(\ref{8}) over all three quarks.

The form (\ref{8}) was used in \cite{16} for the calculation of
pionic transitions in the heavy-light mesons with $g_A^q$ playing
role of a fitting parameter, which turned out to be around 0,7.
This difference from (\ref{9}) can be considered as an indication
of a possible role of nonlocality.

\section{The String Breaking Mechanism}

This mechanism was considered in some detail in \cite{16a} (see
also refs. therein). In the present section we shall consider the
pair creation vertex due to the nonperturbative QCD
configurations. In the $^3P_0$ model \cite{1} this vertex was
modelled by an adjustable constant and in \cite{16a} by some
function $F$. It is our purpose here to derive this vertex from
the basic $4q$ effective action (\ref{2}).

To describe the creation of the $q\bar q$ pair in the presence of
the string which connects quark $Q$ and  antiquark $\bar Q$, one
can  as in \cite{11} take the large $N_c$ limit of the same $4q$
Lagrangian (\ref{2}), which obtains by replacing
$~^f\psi_{b\beta}(x)^g \psi^+_{b\gamma}(y)\to \delta_{fg} N_c
S_{\beta\gamma} (x,y)$.

The resulting Lagrangian describes the creation of the $\bar q q$
pair at the points $x,y$ respectively. Physically it is clear that
$x$ and $y$ should lie on the (deformed) string connecting quark
$Q$ at the point $X$ and antiquark $\bar Q$ at the point $\bar X$.
Since the final two strings connect $X$ and $x$, and $\bar X$, and
$y$, we can also choose the contours of integration along the
strings: e.g. $A_4(x) =\int^X_x du_1 E_1 (u_1) $,  $A_4 (y)
=\int^y_{\bar X} dv_1 E_1 (v_1)$, and the string is along the axis
1. (This change of contours  from (\ref{3}) can be traced to the
effect of cancellation in the sum of contours from the quark $Q$
to the point $Y$ and with opposite sign --from the antiquark $\bar
q$ to the point $Y$, which results in the contour integral between
positions $Q (X)$ and $\bar q(x)$). As a result the kernel $J_E$
is

\be
J_E(\vex, x_4; \vey, y_4) =\int^X_x du_1 \int^y_{\bar X} dv_1
D(u_1-v_1) \cong \frac{\sigma}{\pi} \exp(-\frac{(x_4
-y_4)^2}{4T_g^2}) \label{10} \ee where for $D(x)$ the Gaussian
form was used, c.f. \cite{11}. As in (\ref{7}) one can find the
effective mass operator (due to color electric fields) $M(x,y)$,
using the definition \cite{12},
\be
M(x,y) = -i \gamma_4 S(x,y)\gamma_4 J(x,y) \label{11} \ee and the
estimate of the quark Green's function $S(x,y)$ for a long string,
$L \gg T_g$, done in \cite{11, 14}, gives $S(x,y) \sim i
\delta^{(3)}(x-y)$.

As a result one obtains for the effective Lagrangian the same form
as in (\ref{4}),
\be
  \Delta L^{(SBM)}=i\int d^{4}x d^{4}y\psi^{+}(x) M^{(br)}(x,y)\psi_a(y)
  \label{12}\ee
where
\be
M^{(br)}(x) = \frac{\sigma}{\pi}\delta^{(3)}(\vex -\vey)
e^{-\frac{(x_4 - y_4)^2}{4T_g^2}} \label{13} \ee

Integrating over $d(x_4 - y_4)$ one gets in the Minkowskian
space-time
\be
\Delta L^{(SBM)} = \frac{2T_g \sigma}{\sqrt{\pi}}\int d^4 x
\bar\psi(x) \psi(x) \label{14} \ee

The form (\ref{14}) coincides with that assumed in the $~^3 P_0$
model. At this point one should take into account that the pair in
(\ref{14}) is created at any point $\vex$ in the space, and by the
derivation this point should lie on the world surface of the
string, deforming it from the general minimal area shape to the
minimal area surface passing through the point $\vex, x_4$.

The probability amplitude for such a deformation is equal to
$\exp(-\Delta A(x))$, where $\Delta A(x) = \sigma \Delta S$ is the
increase of the area $\Delta S$. Assuming the latter is described
by the cone with the height $h$ and radius $R$, $h \ll R$, one
obtains an additional factor to be inserted in (\ref{14}), namely,
\be
F^P \equiv \exp(-\Delta A) \simeq \exp (-\sigma\frac{\pi}{2} h^2)
\label{15} \ee

This factor is similar to the one suggested in the $~^3 P_0$
flux-tube model \cite{2}. Integrating over $h$ in (\ref{14})
yields the final result, which we write in the same form as in the
$~^3 P_0$ model \cite{4},
\be
H_I^{(SBM)}=g\int d^3x\bar \psi (x) \psi (x);~~ g=
2T_g\sigma/\sqrt{\pi};~~  \gamma = \frac{T_g
\sigma}{\sqrt{\pi}\mu}=\frac{g}{2\mu}, \label{16} \ee where $\mu$
is the constituent quark mass computed through $\sigma$ \cite{18}.
 Taking $T_g \sim (0.2-0.3)$ fm, $\sigma \approx 0.2$ GeV$^2$ and
$\mu \approx 0.35$ GeV one obtains  $\gamma\sim 0.3-0.5,$ i.e. the
values in the same ballpark as in the phenomenological analysis
\cite{3,4}. Several remarks are now in order. Firstly, above the
simplified Gaussian form of $D(x)$ was used in \cite{10}, which
requires the redefinition of $T_g$; secondly, the contribution of
magnetic correlators and the term $D_1$ in (\ref{1}) has been
neglected. These additional contributions will be considered
elsewhere.

\section{``Perturbative'' pair creation via the hybrid or glueball
formation}

Two decay mechanisms discussed in previous sections are of purely
nonperturbative origin, now we turn to the mechanisms which
contain as a limit a purely perturbative $q\bar q$ pair creation
by a gluon. At large distances one has to know how this process is
modified by the presence of nonperturbative confining fields and
to this end we shall use the Background Perturbation Theory (BPTh)
developed in \cite{10}, where the total gluonic field $A_\mu$ is
separated into valence gluon field $a_\mu$ and background $B_\mu$,
$A_\mu = B_\mu +a_\mu$. The field $B_\mu$ saturates correlator
$D(x)$ and contains therefore its own mass scale, while
perturbation theory is done in powers of $ga_\mu$. The main
physical outcome of the analysis of \cite{10} is that the valence
gluons are propagating inside the film of the string worldsheet,
so that all Feynman diagrams in the coordinate space can be
considered as filled inside by this film on the minimal surface
with boundaries specified by quark and gluon trajectories.

At this point it is clear that one should use the path integral
representation for tha quark and gluon Green's function, namely
the Fock-Feynman-Schwinger (FFS) formalism \cite{19}. The FFS
method has proved useful in conjunction with the BPTh to study
meson, hybrid and glueball Green's functions (see \cite{18} for
review). In \cite{20} this method has been exploited to calculate
mixing between meson, hybrid and glueball states, and in what
follows we shall pursue the same way to study matrix elements of
decays proceeding via hybrid and glueball intermediate states.

We start with the OZI allowed planar pair creation mechanism by a
gluon, propagating inside the film in a hybrid state
$\varphi^{(H)}$, anf therefore the matrix element for the meson
decay via the hybrid states can be written as
\be
W^{(H)} = \sum_n \lambda_n^{(MH)} W_n^{(H)} \label{17} \ee where
$\lambda_n^{(MH)}$ is the dimensionless mixing coefficient of the
$n$-th hybrid state in the given initial meson, which according to
\cite{20} can be written as
\be
\lambda_n^{(MH)} =
\frac{V_{on}^{(\mu)}}{\sqrt{2\mu_g(n)}|M_H^{(n)} -M_M|} \label{18}
\ee

Here $M_H^{(n)}$, $M_M$ are hybrid and meson masses respectively
and $\mu_g(n)$ is the constituent gluon mass in the $n$-th hybrids
state, computed through the string tension as in \cite{18},
\cite{20}.

The matrix element $V_{on}^{(\mu)}$, introduced in \cite{20}, is
that of the pair creation operator $H_1$,
\be
H_1 = g \int \bar q(\vex, 0) \hat a(\vex, 0) q(\vex, 0) d^3 x
\label{19} \ee between the meson state
$\Phi_{\alpha\beta}^{(M)}(\ver)$ and the hybrid state $\Phi_{\mu,
\gamma\delta}^{(H)}(\ver_1, \ver_2)$, where we have specified
quark Dirac indices $\alpha\beta$, $\gamma\delta$ and gluon
component index $\mu$ (this is the 4$\times$4 representation
typical for the Bethe-Salpeter wave functions which is used to
build up $2j+1$ components of meson and hybrid wave functions, for
discussion and refs. see \cite{20}).

In a simplified form $V_{on}$ can be written as \cite{20}
\be
V_{on} = g\int d^3 \ver \phi^{*(H)}(0, \ver)
\Gamma\phi^{(M)}(\ver) d^3 \ver + \mbox{\rm perm.} \label{20} \ee
where "perm" implies another term with gluon emitted by antiquark,
and $\Gamma = \gamma_\mu$.

Similarly the term $W_n^{(H)}$ in (\ref{17}) is the decay
amplitude of the $n$-th hybrid state into two mesons, which can be
written as
\be
W_n^{(H)} = \frac{V_{n, 12}}{\sqrt{2\mu_g(n) V}} \; , \;\; N_{n,
12} \equiv \int \Phi^{*(M_1)}(\ver_1) \Phi^{*(M_2)}(\ver_2)
\Gamma^{(H)}(\ver_1, \ver_2) d^3 \bar r_1 d^3 \bar r_2 \label{21}
\ee

Normalization of wave functions is (summation over repeating Dirac
indices, which are omitted, is implied)
\be
\int|\Phi^{(H)}|^2 d^3 \bar r_1 d^3 \ver_2 = 1 \; , \;\;
\int(\Phi^{(M_i)}|^2 d^3 \ver = 1 \label{22} \ee

Finally two-body decay probability is
\be
w = 2\pi|W^{(H)}|^2 \delta(E_1 - E_2 -M) \frac{Vd^3 k}{(2\pi)^3}
\label{23} \ee

As it is seen from {\ref{17}), (\ref{18}), the decay probability
is strongly enhanced when the decaying meson mass is in the
vicinity of some hybrid level. One should stress that this
situation is standard in the mass range above 1.4. GeV, where the
density of hybrid ground and excited states is fast growing with
mass, (see e.g. second ref. in \cite{6}).

 We now turn to the OZI forbidden, i.e. nonplanar $q\bar
q$ pair creation via valence gluons. The essential step in this
mechanism is the creation of the new flavour quark-antiquark pair
with the new string between them, hence the creation operator
contains at least a two-gluon exchange. In the confining
background trajectories of these two gluons are connected by the
adjoint string (or, equivalently at large $N_c$ by a double
fundamental string) and therefore a new meson is created by the
two- or more gluon glueball, which is emitted virtually from the
original meson. The amplitude for this process can be written as
follows,
\be
W^{(G)} = \sum_n \frac{V_n^{MM_2} V_n^{GM_1}}{M-E_n^G -E(M_2)}
\label{24} \ee where notation is used
\begin{eqnarray}
V_n^{MM_2} & = & \langle\phi^{(M_2)}|\Lambda_n^{(G)}|\phi^{(M)}
\rangle \; , \nonumber \\ V_n^{GM_1} & = & (\phi^{(M_1)*}
\bar\Lambda_n^{(G)}) \label{25}
\end{eqnarray}
and the two-gluon glueball vertex is
\begin{eqnarray}
\Lambda_{n, \alpha\beta}^{(G)}(x_1, x_2) & = & \Psi_{n, \nu_1
\nu_2}^{*(G)}(x_1, x_2) (\gamma_{\nu_1} S^{(f)}(x_1, x_2)
\gamma_{\nu_2})_{\alpha\beta} \nonumber \\
 \bar\Lambda_{n,
\alpha\beta}^{(G)}(y_1, y_2) & = & (\gamma_{\mu_1} S^{(g)}(y_1,
y_2)\gamma_{\mu_2})_{\alpha\beta} \Psi_{n, \mu_1 \mu_2}^{(G)}(y_1,
y_2) \label{26}
\end{eqnarray}
 Here $\Psi^{(G)}$ is the glueball wave function,  and
 $S^{(g,f)}$ - the Green's function of quarks with flavor $(g,f)$.

When all distances $|y_i -x_k|$, $|x_1 - x_2|$, $|y_1 - y_2|$ are
small as compared to $T_g$, the two-gluon Green's function reduces
to the two-gluon exchange of free gluons. This limiting
perturbative mechanism is known for a long time \cite{21}. In the
opposite limit only the lowest mass term survives in the sum
(\ref{24}).

The most interesting case occurs when the denominator in
(\ref{24}) becomes small, which is possible when the mass of some
glueball state is close to the mass of the emitted meson $M_1$.
This amplification may thus occur in the $0^{++}$ channel for
$M_1$ around 1.5 GeV, or in other channels for $M_1\ga 2 $ GeV
\cite{22}.
 A
similar mechanism may take place  in hadron-hadron scattering with
creation of the $c\bar c$ states in the mass range 3 - 4 GeV,
where also glueball states are predicted in lattice and in
analytic calculations \cite{22}.

\section{Conclusion}
 In the present paper the first   qualitative step  was  done
 aimed at the  systematic derivation of strong decay amplitudes
 from the QCD Lagrangian.
We have considered three decay mechanisms, where nonperturbative
contribution is very important.

The first one, the CDM, was already successfully checked in
heavy-light meson decays \cite{16}. The next step would be to
apply CDM to the pionic and kaonic decays of light-light mesons,
and investigate double and more-pion decays, which are given by
the nonlinear Lagrangian (\ref{8}). The second mechanism, the SBM,
turns out to be mostly $~^3 P_0$ (additional terms due to the
smaller nonscalar component in the quark Green's function $S$
considered in \cite{11} have been neglected above) and its
amplitude is close to the phenomenological fits \cite{4}. The
third type of mechanisms with hybrid or glueball in the
intermediate state is the nonperturbative background
generalization of the original purely perturbative mechanisms of
the $~^3 S_1$ type \cite{23} and of two-gluon exchange \cite{21},
respectively. It is argued that a strong enhancement of decay
amplitudes is possible when the corresponding levels of a hybrid
or a glueball are close to the mass of the original or the final
meson respectively. The paper does not contain quantitative
predictions, which are planned for the future, but rather is
concentrated on the general discussion of possible decay
mechanisms as they emerge from the basic QCD Lagrangian.

One of immediate extensions of the present result is the inclusion
of the baryon decays, where all three mechanisms discussed above
are present in the same form with the only replacement of the
simple mesonic string by the $Y$-shaped baryonic string.

Finally, one should have in mind that all three mechanisms
discussed above enter the decay amplitude additively and therefore
one can expect in general some interference effects, which make
the analysis of data more complicated and perhaps more
interesting.

The author is grateful to Yu.S. Kalashnikova and V.I. Shevchenko
for valuable discussions. The financial support of INTAS grants
00-110 and 00-366 and RFBR grants 00-02-17836 and 00-15-96786 is
gratefully acknowledged.

\end{document}